\documentclass[prl,twocolumn,showpacs,preprintnumbers,amsmath,amssymb]{revtex4-1} 

\usepackage{dcolumn}
\usepackage{bm}

\usepackage{amsmath}
\usepackage{amssymb}
\usepackage{amsfonts} 
\usepackage{wasysym}  
\usepackage{graphics}  
\usepackage{psfrag} 
\usepackage{graphicx} 
\usepackage{epsfig}    
\usepackage{rotating}  
 \usepackage[latin1]{inputenc}
 \usepackage{color}
 \usepackage{rotating}
 \usepackage{multirow}

\usepackage{color}
\usepackage{epsfig}
\definecolor{darkgreen}{rgb}{0,0.5,0} 
\definecolor{violet}{rgb}{0.5,0,0.5}
\definecolor{orange}{rgb}{0.2,0.5,0.5}
 
\newcommand{\bulk}[0]{\text{b}} 
\newcommand{\res}[0]{-} 

\begin{document}


\title{Microtubule Length-Regulation by Molecular Motors}

%
\author{Anna Melbinger}
\email{These authors contributed equally to this work}
\author{Louis Reese}
\email{These authors contributed equally to this work}
\author{Erwin Frey}
\email{frey@lmu.de}

\affiliation{
Arnold Sommerfeld Center for Theoretical Physics (ASC) and Center for NanoScience (CeNS), Department of Physics, Ludwig-Maximilians-Universit\"at M\"unchen, Theresienstrasse 37, 80333 M\"unchen, Germany}

\begin{abstract}
Length-regulation of microtubules (MTs) is essential for many cellular processes.  Molecular motors like kinesin 8, which move along MTs and also act as depolymerases, are known as key players in MT dynamics. However, the regulatory mechanisms of length control remain elusive. Here, we investigate a stochastic model accounting for the interplay between polymerization kinetics and motor-induced depolymerization. We determine the dependence of MT length and variance on rate constants and motor concentration. Moreover, our analyses reveal how collective phenomena lead to a well-defined MT length.
\end{abstract}



\maketitle

During the lifespan of an eukaryotic cell microtubules (MTs) perform highly dynamic tasks. For instance, during mitosis, they form the mitotic spindle, which searches, captures, and separates the double set of chromosomes~\cite{Alberts2002}.  To achieve such complex dynamic behavior there need to be molecular mechanisms which allow a dynamic control of MT length. There is much evidence that these mechanisms rely on an intricate interplay of GTP hydrolysis~\cite{Mitchison1984}, mechanical forces~\cite{Goshima2010,*Dumont2009}, and regulatory proteins~\cite{Wordeman2005,*Howard2007}.  In particular, the role of the molecular motor families  kinesin-5 and kinesin-8 has been investigated: Several \emph{in vivo} experiments showed that both, the presence and the concentration of such proteins, strongly affect the functionality of the mitotic spindle~\cite{Goshima2005,Gupta2006,*Mayr2007,*Stumpff2008,*Tischer2009}. This is supported by \emph{in vitro} experiments which specifically studied the molecular mechanisms of interactions between motor proteins and microtubules~\cite{Varga2006,Varga2009,Gardner2011a,Du2010,Su2011,Weaver2011,Bieling2010}. In general, it is accepted that kinesin-8 hampers MT growth. In particular, it was found  that the plus-end directed motor kinesin-8 of budding yeast, Kip3p, depolymerizes MTs at the tip. To gain a deeper understanding for the molecular mechanisms underlying these depolymerization dynamics Varga et al.~\cite{Varga2006,Varga2009}  studied the interaction of Kip3p with stabilized MTs not exhibiting dynamic instability~\cite{Mitchison1984,Dogterom1993}. The key result of these experiments is that depolymerization is length-dependent,~\emph{i.e.}, longer MTs depolymerize faster than shorter ones. One main determinant of the observed length-dependence are molecular traffic jams which can successfully be described by driven diffusive processes~\cite{Reese2011}. These findings suggest, that length-dependent depolymerization in combination with polymerization allows a cell to regulate the length of MTs~\cite{Varga2006,Varga2009}. There are by now several theoretical studies addressing length-regulation ranging from MTs \cite{Govindan2008,Tischer2010}, over actin filaments~\cite{Erlenkamper2009} to fungi~\cite{Sugden2007,*Sugden2007a} and flagellae~\cite{Schmitt2011}.

In this Letter, we study the combined influence of spontaneous MT polymerization and motor induced depolymerization.  In our model we neglect MT dynamics at the minus end as there the dynamics rates are much smaller than the ones at the plus end~\cite{Alberts2002}. Furthermore, under physiological conditions often the minus end dynamics are completely suppressed due to capping proteins~\cite{Desai1997}. We build on a recently validated quantitative model for MT depolymerization~\cite{Varga2009,Reese2011}, and extend it by introducing polymerization dynamics at the fast-growing plus-end~\cite{Alberts2002}.
This accounts for MT growth mediated by spontaneous~\cite{Mitchison1984} or enzymatically catalyzed~\cite{Brouhard2008} attachment of tubulin heterodimers to the tip. This approach enables us to study the basic principles underlying length-regulation which is achieved by the antagonism between length-dependent depolymerization and spontaneous polymerization dynamics. We predict quantitative criteria for the parameter regime where regulation is feasible. In addition, we calculate both the mean length and the corresponding standard deviation, and thereby determine the accuracy at which regulation is achieved. 
\begin{figure}[!t]
\centering
\includegraphics[width=0.98\columnwidth]{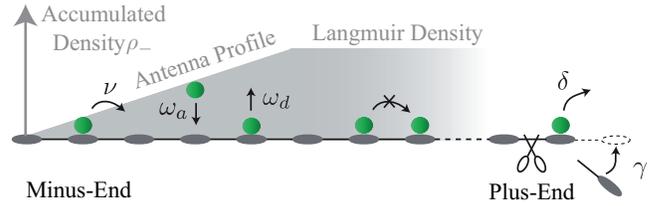}
\caption{Illustration of the model. Motors attach to and detach from the MT lattice at rates $\omega_a\!=\!c\tilde{\omega}_a$ and $\omega_d$, respectively. On the lattice particles hop to the right at rate $\nu$ provided that the next site is empty. At the right boundary, the MT plus-end, particles remove the last lattice site at rate $\delta$ and the MT lattice grows at rate $\gamma$. The resulting antenna-like density profile $\rho_-(x)$ is sketched in light gray. 
\label{fig:cartoon}}
\end{figure}
To describe the MT dynamics we employ a driven diffusive lattice gas model~\cite{Lipowsky2001,*Klumpp2003,Parmeggiani2003,*Parmeggiani2004} as illustrated in Fig.~\ref{fig:cartoon}.  Since MT protofilaments serve as independent tracks for the motors~\cite{Ray1993,Vilfan2001}, a MT can effectively be described by a one-dimensional lattice of dynamic size $L(t)$. The size of a tubulin heterodimer sets the basic length scale of the lattice. The state of each site, $i$, is described by its occupation number, $n_i\!\in\!\{0,1\}$, where $n_i\!=\!0$ and $n_i\!=\!1$ signify an empty and occupied site, respectively. On the MT lattice the dynamics follow the \emph{totally asymmetric simple exclusion process} with Langmuir kinetics~\cite{Parmeggiani2003}: Motors can attach to and detach from the MT at rates $\omega_a\!=\!c\tilde{\omega}_a$ and $\omega_d$, respectively, where $c$ is the motor concentration in the surrounding fluid; the binding constant is defined as $K\!:=\!\omega_a\!/\!{\omega}_d$. On the lattice, particles move to right at rate $\nu$ provided that the next site is empty; $\nu\!=\!1$ sets the basic time scale.  The combined effect of motor attachment in proximity of the minus-end and subsequent movement towards the plus-end leads to an accumulation of motors, which finally results in an antenna-like steady state profile~\cite{Varga2009,Hough2009,Reese2011} as illustrated in Fig.~\ref{fig:cartoon}.  At a certain distance from the minus-end the density profiles saturate to the equilibrium Langmuir density $\rho_\text{La}=K/(K+1)$~\cite{Leduc2012}. The resulting accumulated density profiles in vicinity of the minus-end, $\rho_-(x)$, can be described by Lambert-$W$ functions~\cite{Parmeggiani2004,Reese2011}. Moving further towards the right boundary (MT minus-end), the density profile is determined by the interplay of motor current and the boundary conditions at the plus-end. This entails a rich variety of collective phenomena and leads to nontrivial density profiles~\cite{Schuetz2000,Chou2011}. In the present study, the right \emph{boundary is dynamic}. Motivated by the recent  studies on kinesin-8~\cite{Varga2006,Varga2009,Gupta2006,Gardner2011a,Du2010,Su2011,Weaver2011}, we  consider the following scenario: When a motor arrives at the MT tip, it detaches by removing the last MT-site at rate $\delta$~\cite{Reese2011}. In addition, subsuming the effects of spontaneous and enzymatic polymerization, the MT is assumed to polymerize through the attachment of single tubulin heterodimers at an effective rate $\gamma$. These boundary conditions lead to a dynamic MT length which is determined by the combined effect of the particle current onto the last site, polymerization, and depolymerization rates. 


The dynamic length of the MT, $L(t)$, is determined by the particle density at the MT plus-end $\rho_+(L)$,
\begin{equation}
\partial_t L (t)=-\delta\rho_+(L)\!+\!\gamma .
\label{eq:dtL}
\end{equation}
This equation defines a critical density $\rho_+^c\!=\!{\gamma}/{\delta}$, at which the MT length is in a steady state, $\partial_t L\!=\!0$. For tip densities smaller or larger than $\rho_+^c$ the MT grows or shrinks, respectively. As the tip density is fed by the motor current towards the tip, it depends on the accumulated motor density in bulk $\rho_-(x)$. This suggests the following mechanism for MT length-regulation: On short MTs, the accumulated motor density is low, and therefore also the tip-density $\rho_+(L)$. As long as $\rho_+(L)\!<\!\rho_+^c$ the MT grows. In contrast, for longer MTs higher accumulated motor and tip densities are reached which eventually result in MT depolymerization once $\rho_+(L)\!>\!\rho_+^c$. However, this mechanism is only expected to work if the tip is not growing too fast: Above a critical polymerization rate the particle current feeding the tip density can no longer follow the advancing tip.

To quantify these heuristic arguments and determine the precise conditions under which length-regulation is feasible and which length is adjusted, the tip density has to be determined. This requires to analyze the intricate interplay between molecular crowding due to high motor density~\cite{Lipowsky2001,*Klumpp2003,Parmeggiani2003,*Parmeggiani2004}  and transport bottlenecks at the plus-end~\cite{Pierobon2006,Reese2011}. In addition, this boundary is highly dynamic, and  calculations of the tip density are more intricate than for standard driven diffusive models for which the size of the lattice is constant~\cite{Schuetz2000,Nowak2007, Chou2011}.

\begin{figure}
\centering
\includegraphics[width=\columnwidth]{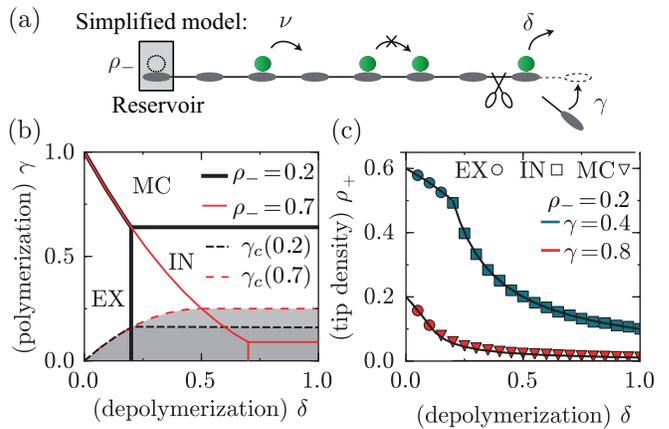}
\caption{(a) Illustration of the SM. 
(b)~Phase diagram as a function of the depolymerization and polymerization rates, $\delta$ and $\gamma$, respectively. The gray shaded area indicates regions in phase space in which regulation is possible in the full model. The gray area indicates regions in phase space where the MT shrinks in the SM. In the full model, for $\rho_-=\rho_\text{La}$ length regulation is only possible in the gray area as detailedly explained in the main text. (c)~Comparison of simulation data with analytical results for the tip density $\rho_+$.
\label{fig:PD}}
\end{figure}

To make further progress, we first consider a simplified model (SM) where we disregard spatial variations of the density profile. In detail, we assume a constant density $\rho_\res$ that serves as a particle reservoir at the left boundary, neglect attachment and detachment kinetics, but leave the dynamics at the plus-end unchanged, see Fig.~\ref{fig:PD}(a). This allows us to focus on the dynamics at the plus-end and to unravel how they depend on the reservoir density $\rho_\res$. Since, we find that the density profiles adapt adiabatically to a dynamic lattice size~\footnote{This assumption is valid for tip dynamics slower than the motor speed, which is true in the parameter regime where length-regulation is possible.}, the results for the full model can  be inferred upon replacing $\rho_\res$ by $\rho_\res (x)$. As the length of the lattice is dynamic, we perform our calculations in a comoving frame fixed to the right boundary. In this frame, a polymerization event corresponds to the simultaneous movement of all particles on the lattice to the minus-end by one unit, while depolymerization results in an instantaneous shift to the right. 
Thus, in a mean-field approximation [$\langle n_in_j\rangle=\langle n_i\rangle\langle n_j\rangle= \rho_i\rho_j$] the particle current in bulk is given by,
\begin{align}
J(\rho_\bulk,\rho_\text{+})=\rho_\bulk(1-\rho_\bulk)-\gamma\rho_\bulk+\delta\rho_+\rho_\bulk,
\end{align}
where $\rho_\bulk$ is the motor density in bulk. The first term describes the hopping processes, the second and third term account for simultaneous movement of all particles due to polymerization and depolymerization, respectively. Importantly, the bulk current explicitly depends on the tip density and thereby on the right boundary.


To determine the phase behavior we employ the \emph{Extremal Current Principle} (ECP)~\cite{Krug1991,Kolomeisky1998,Popkov1999} relying on two velocities: The \emph{collective velocity} $v_\text{coll}(\rho)\!=\!\partial_\rho J$ determines the direction in which a local density perturbation spreads. Thereby, one is able to determine whether a certain  bulk density is stable against perturbations, \emph{i.e.} for a density $\rho$ stable at the left (right) boundary $v_\text{coll}(\rho)\!>\!0$ ($v_\text{coll}(\rho)\!<\!0$) holds. The boundary conditions result in densities at the plus and the minus-end, respectively, whose stabilities can now be tested employing $v_\text{coll}$. If these densities are stable against small perturbations, we call them $\rho^{\text{left}}$ and $\rho^{\text{right}}$ as they are given by the system's left and right  boundary, respectively. If either one or both of these boundary densities are not stable, perturbations change these densities and  $\rho^{\text{left}}$ and $\rho^{\text{right}}$ are given by the first stable density which is determined by  $v_\text{coll}(\rho)=0$. The \emph{shock velocity} $v_\text{shock}(\rho^{\text{left}},\rho^{\text{right}})\!=\!(J(\rho^{\text{left}})\!-\!J(\rho^{\text{right}}))/(\rho^{\text{left}}\!-\!\rho^{\text{right}})$ determines the direction in which a virtual domain wall between the densities at the left and the right, $\rho^{\text{left}}$ and $\rho^{\text{right}}$, moves and thereby which of these densities is realized in bulk. In more detail, for $v_\text{shock}>0$ the left density, $\rho^\text{left}$, dictates the bulk density, while for $v_\text{shock}<0$ the right density, $\rho^\text{right}$, is realized.
In our model particles are transported to the right and therefore jams spread from right to left. Hence, the virtual domain wall arises at the right boundary and  the tip densities $\rho_+^\text{left}$ and $\rho_+^\text{right}$ determine $v_\text{shock}$; see Supporting Material. 


Due to particle conservation at the plus-end of the MT, $\partial_t\rho_+=J(\rho_\bulk,\rho_+)-\rho_+\delta$, the stationary value of the bulk and tip density are related through
\begin{equation}
\rho_+\delta = J(\rho_\bulk,\rho_+) \, .
\label{eq:tipdensity}
\end{equation}
This implies that the values of these densities as well as the nature of the ensuing non-equilibrium steady state are strongly interlinked. In general, one expects three phases~\cite{Krug1991,Derrida1992,*Schuetz1993,*Derrida1993}: the steady state may either be dominated by the motor densities at the plus-end (EX-phase) and the minus-end (IN-phase), respectively, or by the transport capacity (maximal current) of the lattice itself (MC-phase). We first consider the IN-phase where $\rho_\bulk^\text{IN}\!=\! \rho_\res$ holds, and Eq.~\eqref{eq:tipdensity} leads to the tip density 
\begin{equation}
\rho_{+}^\text{IN}(\rho_\res)=\rho_\res(1\!-\!\rho_\res\!-\!\gamma)/[\delta(1\!-\!\rho_\res)] \, . 
\label{eq:tip_density_IN}
\end{equation}
This solution is stable against perturbations only if the collective velocity $v_\text{coll} (\rho_\res)\!=\!\partial_\rho J(\rho,\rho_{+}^\text{IN}(\rho)) |_{\rho=\rho_\res}$ is positive, which holds for reservoir densities smaller than the bulk density in the MC-phase $\rho_\bulk^\text{MC}\!=\!1\!-\!\sqrt{\gamma}$. If the reservoir density exceeds this value, the ECP implies that the tip density becomes constant and independent of the reservoir density $\rho_{+}^\text{MC}\!=\!(1\!-\!\sqrt{\gamma})^2/\delta$. 
For the EX-phase, the right boundary determines the bulk density $\rho_\bulk^\text{EX}\!=\!\rho_+^\text{EX}$, and  Eq.~\eqref{eq:tipdensity} leads to $\rho_{+}^\text{EX}\!=1\!-\gamma/(1-\delta)$. 
According to the ECP, this solution is stable if the corresponding collective velocity $v_\text{coll} (\rho_{+}^\text{EX})\!=\!\partial_{\rho}J(\rho,\rho_{+}^\text{EX})|_{\rho=\rho_{+}^\text{EX}}$ is negative. Since in the relevant parameter regime $v_\text{coll}\!<\!0$ is always fulfilled, the density $\rho_{+}^\text{EX}$ is always stable and $\rho_+^\text{right}\!=\!\rho_+^\text{EX}$ holds.

\begin{figure*}[!t]
\centering
\includegraphics[width=2\columnwidth]{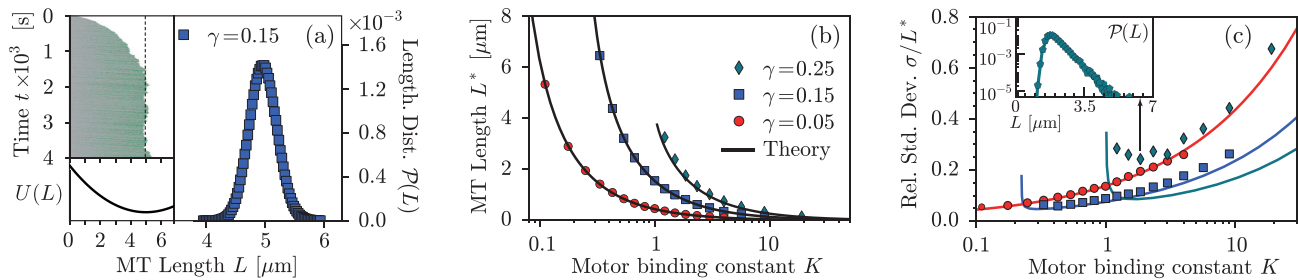}
\caption{(a)~Kymograph (upper left): Molecular motors (green shaded traces) accumulate along the lattice (gray) resulting in a steady MT length (dashed). Corresponding potential $U(L)$ (bottom left) and length distribution $\mathcal{P}(L)$ (right) for $K=1.5$.
(b)~Analytical (lines, Eq.~\eqref{eq:typic}) and numerical  results (symbols) for the typical MT length $L^*$ are compared.
(c)~Standard deviation of the MT length $\sigma$ in units of its typical length $L^*$ for the same values for $\gamma$ as in (b). Inset: $\mathcal{P}(L)$ for $\gamma\!=\!1/4$ shows an exponential tail. [Parameters: $\delta\!=\!0.5$ $\gamma_c=0.25$ (a-c), $\omega_d\!=\!2\!\times\!10^{-4}$ (a), and $\omega_d\!=\!1\!\times\!10^{-3}$ (b), (c)].
\label{fig:stat}\label{fig:ruler}}
\end{figure*}

In summary, we have found the following results for the densities at the left and right boundary of the MT:
\begin{eqnarray}
 \rho_{+}^{\text{left}}       = \text{Min}[\rho_{+}^\text{IN},\rho_{+}^\text{MC}] \, , \qquad
 \rho_{+}^{\text{right}}     = \rho_{+}^\text{EX} \, .
\end{eqnarray}
With these expressions at hand, we can now map out the phase diagram upon evaluating the shock velocity $v_\text{shock}(\rho_+^\text{left},\rho_+^\text{right})$, cf. Fig.~\ref{fig:PD}. The IN-phase is determined by $\gamma<(1-\rho_\res)^2$ and $\delta>\rho_\res$. Importantly, it is the only phase in which the tip density is a function of $\rho_\res$; see Eq.~\eqref{eq:tip_density_IN}. As $\rho_\res$ corresponds to the spatially varying density profile $\rho_\res (x)$  in the full model, length-regulation is feasible in this range of parameters. In contrast, in the EX-phase [$\gamma<(1-\delta)^2$ and $\delta<\rho_\res$] and the MC-phase [$\gamma>(1-\delta)^2$ and $\gamma>(1-\rho_\res)^2$] neither the tip nor the bulk densities depend on $\rho_\res$. To confirm these and the following analytic results, we performed extensive stochastic simulations  employing the Gillespie algorithm~\cite{Gillespie1977,*Gillespie1976}.  For both, the simplified  and the full model discussed in the following, calculations are in excellent agreement with simulations, cf. Fig.~\ref{fig:PD}(c) and Fig.~\ref{fig:ruler}.

Moreover, upon combining the results for the tip densities in the various phases with the critical density $\rho_+^c=\gamma/\delta$, we are able to calculate the critical growth rate $\gamma_c$, at which the MT length becomes stationary:
\begin{equation}
\gamma_c(\rho_-)=\begin{cases} \delta (1-\delta) &\text{EX-phase} \, , \\
\rho_\res(1-\rho_\res)  &\text{IN-phase}  \, , \\
1/4  &\text{MC-phase} \, .
\end{cases}
\end{equation}
For $\gamma>\gamma_c$ the lattice grows to infinity, while it shrinks indefinitely for $\gamma<\gamma_c$.

Up to now, the discussion was restricted to a simplified system, and we have learned how a constant reservoir density translates into the tip density and in which parameter regimes the MT grows and shrinks, respectively. In the following we transfer the so far obtained results to the full spatial model, in which the reservoir density is replaced by the density profile: $\rho_\res \to \rho_-(x)$. This implies that  also the tip density becomes length-dependent, $\rho_+\to\rho_+(L)$ in the IN-phase, see Eq.~\eqref{eq:tip_density_IN}. Let us first consider how  these spatial density profiles affect the critical growth rate, $\gamma_c$, in the full model, and thereby derive a condition for the parameter regime where length-regulation is feasible: Growth is unbounded only if the highest accumulated density $\rho_\res(L)$ does not result in strong enough depolymerization dynamics to overcome MT growth due to polymerization.  Recall that the accumulated density profile increases from left to right until it saturates to the Langmuir density, $\rho_\text{La}$. Thus, growth is unbounded if $\gamma \!>\! \gamma_c ({\rho_\text{La}})$. In contrast, in the depolymerizing regime, $\gamma \! < \! \gamma_{c}({\rho_\text{La}})$, the MT shortens until the tip enters the antenna profile. Within this regime, the accumulated density and thereby the tip density decrease with every depolymerization event until the MT length reaches a stable fixed point $L^*$, at which growth and shrinkage balance each other. As the corresponding restoring force is conservative, the length regulation dynamics can be described by a  potential $U$. It follows from $-\partial_L U=-\delta\rho_+(L)\!+\!\gamma$ and leads to an adjusted length fluctuating around the mean, as observed in  the MT dynamics; see Fig.~\ref{fig:stat}(a).

To calculate the adjusted MT length $L^*$, the full spatial density profile $\rho_\res (x)$ as obtained from mean-field theory~\cite{Parmeggiani2003}, and the stochastic growth and shrinkage have to be considered. They can be combined in an effective master equation, where the degrees of freedom from the occupation numbers, $n_i$, are adiabatically eliminated:
\begin{equation}
\partial_t \mathcal{P}(L)\!=\! \left [ (\mathbb{E}^+\!-\!1)\delta \rho_+(L)\!+\!(\mathbb{E}^-\!-\!1) \gamma \right ] \mathcal{P}(L).
\label{eq:me}
\end{equation}
Here, $\mathbb{E}^\pm$ are step operators which increase or decrease the lattice length; $\rho_+(L)$ is the density at the tip depending on $L$. In the IN-phase, in which regulation is feasible, $\rho_+(L)\!=\!\rho_\res(L)(1-\rho_\res (L)-\gamma)/[\delta(1-\rho_\res (L))]$ holds, where $\rho_\res (x)$ is the spatial density profile given by Lambert-$W$ functions~\cite{Parmeggiani2003,Reese2011}. We solve the Master equation approximately using the 
van Kampen system size expansion~\cite{VanKampen2007}: The deterministic dynamics  $\ell(t)$ is separated from the fluctuations $\xi$ employing the ansatz $L=\Omega\ell(t)+\sqrt{\Omega}\xi$. As expansion parameter we consider $\Omega\!=\!1/\omega_a$ because the typical length scale of the accumulated density profile which triggers length-regulation is given by $1/\omega_a$. Additionally, time has to be rescaled according to $\tau\!=\!\omega_a t$ since the equilibration time also scales with this length scale. An expansion of Eq.~\eqref{eq:me} in terms of $1/\sqrt{\Omega}$ yields the mean MT length
\begin{equation}
L^*\!=\!\tfrac{\rho_\text{La}}{\omega_a}  \left (1\!-\!\sqrt{1\!-\!4\gamma}\!+\! \tfrac{K-1}{K+1} \ln|\tfrac{(K+1)\sqrt{1-4\gamma}+ K -1}{2 K}|\right) \, .
\label{eq:typic}
\end{equation}
As can be inferred from  Fig.~\ref{fig:stat}(b), this result is in excellent agreement with numerical data. We observe that the stationary MT length is independent of $\delta$, and a monotonically decreasing function of the binding constant $K$. The latter behavior reflects the increase of the slope of the antenna profile with larger $K$ implying that the density at which regulation arises is reached for shorter MTs. The van Kampen approximation also gives the variance,
\begin{equation}
\sigma^2\!=\!\frac{2 \gamma^2}{\omega_a} \,  \frac{K}{-1+\sqrt{1-4\gamma} +2\gamma(1+K)}\label{eq:sigma}.
\end{equation}
For small values of  $\gamma$, the standard deviation $\sigma$ is below $10 \%$ of the filament length in a range of approximately $1\,\mu \text{m}\!-\!20\,\mu \text{m}$. The variance actually diverges with $\gamma\! \to\! \rho_\text{La} (1\! - \!\rho_\text{La})$ for $K\leq1$, while for $K>1$ regulation remains possible for $\gamma=\rho_\text{La} (1\! -\! \rho_\text{La})$. In this regime, the MT length distribution $\mathcal{P}(L)$ develops an exponential tail. This tail cannot be described by
the van Kampen expansion, which explains the deviations between the numerical and the analytical results  in Fig~\ref{fig:stat}(c).

In this Letter, we investigated how motor-induced depolymerization in combination with spontaneous polymerization can result in length-regulation of biological filaments. We found a broad parameter regime in which length-regulation is feasible, due to collective phenomena of molecular motors which also act as depolymerases.  Even though the regime where length-regulation is possible depends on the depolymerization rate, the  adjusted filament length is independent of the depolymerization rate $\delta$, because of microscopic traffic jams forming at the tip. Our model provides a proof of principle that spatial dependences in the growth and shrinkage rates of filaments, which arise from motor transport in this case, can result in a well-defined filament length. It may serve as a basis for mechanistically more detailed analyses which account for multiple protein species~\cite{Ebbinghaus2010}, dynamic instability~\cite{Brun2009,Padinhateeri2012}, internal states of MTs or motors~\cite{Nishinari2005}, assemblies of MTs~\cite{Campas2008}, or the abundance of molecules in the cell~\cite{Brackley2010}. We expect, however, that the main idea - feedback between polymerization dynamics and collective motor dynamics - remains the core mechanism.
\

\begin{acknowledgments}
This project was supported by the Deutsche Forschungsgemeinschaft in the framework of the SFB 863 and the German Excellence Initiative via the program ``Nanosystems Initiative Munich'' (NIM).
\end{acknowledgments}

\bibliographystyle{apsrev4-1}

\newpage
\onecolumngrid

\setcounter{page}{1}

\begin{center}
{\bf \large Supporting Material}\\
\vspace*{0.5cm}
{\large Anna Melbinger, Louis Reese, and Erwin Frey}\\
\vspace*{0.5cm}
{\large Supplementary EPAPS document}
\end{center}
In this supporting information, we give an intuitive argument for the \emph{extremal current principle} to provide some background which facilitates understanding.  Furthermore, we explicitely show, why the density at the right boundary $\rho_{+}^\text{EX}$ is always stable in the relevant parameter regime. 

\section{Phase Behavior}
\subsection{An Intuitive Argument for the Extremal Current Principle}

\begin{figure}[b]
\centering
\includegraphics[width=15cm]{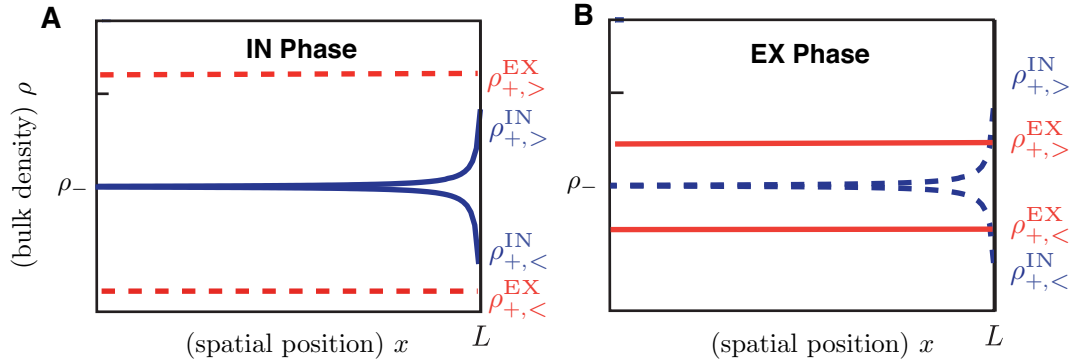}
\caption{Illustration of the right boundary determining the phase behavior for bulk densities, $\rho_-$.  Density profiles corresponding to the IN-phase are shown in blue while the ones corresponding to the EX-phase are red. The tip densities can either both be smaller or larger than the bulk density,  $\rho_{+,<}^\text{IN}$, $\rho_{+,<}^{\text{EX}}$ and $\rho_{+,>}^\text{IN}$, $\rho_{+,>}^\text{EX}$, respectively. Both scenarios are depicted here. If the dynamics are determined by the entering current as in ({\bf A}), the tip density, which would arise in the EX-phase, $\rho_{+}^{\text{EX}}$, serves as a virtual bottleneck. Therefore, the tip density corresponding to the maximal exiting current is realized. In other words, the IN-phase arises for $\rho_{+,<}^{\text{IN}}>\rho_{+,<}^\text{EX}$ or $\rho_{+,>}^{\text{IN}}<\rho_{+,>}^{\text{EX}}$ ({\bf A}), while the EX-phase emerges for  $\rho_{+,<}^\text{IN}<\rho_{+,<}^\text{EX}$ or $\rho_{+,>}^\text{IN}>\rho_{+,>}^\text{EX}$ ({\bf B}).\label{fig:Ill_ECP}}
\end{figure}

To decide which phase is realized in bulk the \emph{extremal current principle} (ECP) can be used~\cite{Krug1991a,Kolomeisky1998a,Popkov1999a,Hager2001a}. Employing the shock velocity, $v_\text{shock}$ one can derive whether the right or the left boundary determines the phase behavior. This means that the  phase transition lines between IN- and EX-phase or MC- and EX phase can be calculated. Note, that the second order transition between IN- and MC-phase is due to the collective velocity (as discussed in the main text) and therefore the description employed here does not apply. In the following, we provide an intuitive argument based on traffic jams on the lattice, to decide which phase is realized. As it will turn out, this leads to the same result as obtained from the shock velocity. An additional merit of the heuristic arguments is that they provide insights into why it is necessary to consider the tip densities instead of the bulk densities in the shock velocity, $v_\text{shock}(\rho_+^\text{left},\rho_+^\text{right})$. For simplicity, let us assume $\rho_-<\rho_\text{max}$, \emph{i.e.} the maximal current never determines the transport in bulk. Thus, we focus on the IN- and EX-phase and the corresponding phase transition. The regime $\rho_->\rho_\text{max}$ and the phase transition from MC- to EX-phase can be analyzed analogously simply by replacing $\rho_-$ by $\rho_\text{max}$. For  $\rho_-<\rho_\text{max}$, there are two possible scenarios for the bulk densities as sketched in Fig.~\ref{fig:Ill_ECP}: In the IN-phase (blue line), the density at the left boundary is given by $\rho_-$  and the tip density has a distinct value, $\rho_+^\text{IN}$, which lies either above or below $\rho_-$. In contrast, in the EX-phase (red line), both the bulk and the tip density are given by the same value $\rho_+^\text{EX}$. Before turning to the question which phase is realized depending on the parameters, we first show that either both possible tip densities lie above $\rho_-$ (indicated by a '$>$' subscript), $\rho_{+,>}^{\text{IN}}$ and $\rho_{+,>}^\text{EX}$, or below $\rho_-$ (indicated by a '$<$' subscript), $\rho_{+,<}^\text{IN}$ and $\rho_{+,<}^\text{EX}$.  Upon employing Eq.~(4) from the main text for the tip density, $\rho_-\gtrless\rho_+^\text{IN}$ can be expressed as,
\begin{align}
\rho_-\gtrless\frac{\rho_-(1-\rho_--\gamma)}{\delta(1-\rho_-)}.
\end{align}
Rearraging yields,
\begin{align}
\rho_-\gtrless1-\frac{\gamma}{1-\delta}=\rho_+^\text{EX}.
\end{align}
With these results at hand we can now decide which phase is realized in the system. To this end we need to compare the current from the left [main text Eq.~(1), $J^\text{left}=J(\rho_-,\rho_+^\text{IN})$] and the current from the right [$J^\text{right}=J(\rho_+^\text{EX},\rho_+^\text{EX})$]. If  $\rho_{+,<}^\text{IN}>\rho_{+,<}^\text{EX}$ or $\rho_{+,>}^\text{IN}<\rho_{+,>}^\text{EX}$ holds, the particle current from the left, which depends on $\rho_+^\text{IN}$ [main text Eq.~(1), $J(\rho_-,\rho_+^\text{IN})$], is always smaller than the one from the right, which depends on $\rho_+^\text{EX}$ [$J(\rho_+^\text{EX},\rho_+^\text{EX})$]. Therefore, no traffic jams arise at the right boundary. Therefore the density profile resulting from the left boundary is not disturbed and the IN phase is present, see Fig.~\ref{fig:Ill_ECP}{\bf A}. In contrast, for  $\rho_{+,<}^\text{IN}<\rho_{+,<}^\text{EX}$ or $\rho_{+,>}^\text{IN}>\rho_{+,>}^\text{EX}$, the tip density determined by the exiting current, $\rho_{+}^\text{EX}$, acts as a bottleneck and a traffic jam results at the tip. As this traffic jam becomes macroscopic, i.e. it spreads back into the system it results in a bulk density given by the tip density $\rho_{+}^\text{EX}$; the system is in the EX phase (Fig~\ref{fig:Ill_ECP}{\bf B}). Analogously, one can analyze the phase transition between the MC- and EX-phase for Langmuir densities larger than $\rho_\text{max}$.

\section*{Stability of the right boundary density $\rho^\text{right}_+$}
In analogy to the treatment of the left boundary in the main text, we here analyze the collective velocity at the right boundary  to decide whether $\rho_+^{\text{EX}}$ is stable or unstable. The collective velocity is given by
\begin{equation}
v_c (\rho_{+}^\text{EX})\!=\!\partial_{\rho}J(\rho,\rho_{+}^\text{EX})|_{\rho=\rho_{+}^\text{EX}}=1-2\rho_{+}^\text{EX}-\gamma+\delta\rho_{+}^\text{EX}.
\end{equation}
Thus the $\rho_{+}^\text{EX}$ is stable (unstable) if $\rho_{+}^\text{EX}>\frac{1-\gamma}{2-\delta}$ ($\rho_{+}^\text{EX}<\frac{1-\gamma}{2-\delta}$) holds.
Equating this condition with $\rho_{+}^\text{EX}=1-\gamma/(1-\delta)$, one calculates in which parameter regime the density at the right boundary is stable. Thereby one arrives at the condition,
\begin{equation}
\gamma \leq(1-\delta)^2.
\end{equation}
This is identical to the boundary between the EX and the MC phase. Therefore, the parameter regime where the $\rho_{+}^\text{EX}$ is unstable is not important as the $\rho^\text{left}_+$ then determines the system.

\bibliography{collection1}
\bibliographystyle{apsrev4-1}

merlin.mbs apsrev4-1.bst 2010-07-25 4.21a (PWD, AO, DPC) hacked

\end{document}